\documentclass[fleqn,twoside]{article}
\usepackage{epsf}

\def\onecol{\onecolumn \mathindent 2em}
\def\twocol{\twocolumn \mathindent 1em}
\newcommand{\bls}[1]{\renewcommand{\baselinestretch}{#1}}
\def\noi{\noindent}

\makeatletter
\renewcommand{\section}{\@startsection{section}{1}{0pt}%
        {-3.5ex plus -1ex minus -.2ex}{2.3ex plus .2ex}%
        {\large\bf\protect\raggedright}}

\renewcommand{\subsection}{\@startsection{subsection}{2}{0pt}%
        {-3ex plus -1ex minus -.2ex}{1.4ex plus .2ex}%
        {\normalsize\bf\protect\raggedright}}

\renewcommand{\thesubsubsection}%
        {\arabic{section}.\arabic{subsection}.\arabic{subsubsection}.}

\renewcommand{\@oddhead}{\raisebox{0pt}[\headheight][0pt]{%
   \vbox{\hbox to\textwidth{\rightmark \hfil \rm \thepage \strut}\hrule}}}
\renewcommand{\@evenhead}{\raisebox{0pt}[\headheight][0pt]{%
   \vbox{\hbox to\textwidth{\thepage \hfil \leftmark \strut}\hrule}}}
\newcommand{\heads}[2]{\markboth{\protect\small\it #1}{\protect\small\it #2}}
\newcommand{\Acknow}[1]{\subsection*{Acknowledgement} #1}
\makeatother

\newcommand{\Title}[1]{\noi {\Large #1} \\}

\newcommand{\Authors}[4]{\noi
        {\large\bf #1\dag\ #2\ddag}\medskip\begin{description}
        \item[\dag]{\it #3} \item[\ddag]{\it #4}\end{description}}
\newcommand{\Abstract}[1]{\vskip 2mm \begin{center}
        \parbox{16.4cm}{\small\noi #1} \end{center}\medskip}

\newcommand{\foom}[1]{\protect\footnotemark[#1]}

\newcommand{\email}[2]{\footnotetext[#1]{e-mail: #2}}

\newcommand{\sect}[1]{Sec.\,#1}

\def\nqq{\hspace*{-2em}}
\def\nhq{\hspace*{-0.5em}}

\def\cm{\hspace*{1cm}}
\def\inch{\hspace*{1in}}

\def\ten#1{\mbox{$\cdot 10^{#1}$}}
\def\deg{\mbox{${}^\circ$}}                     

\def\al{&\nhq}
\def\lal{&&\nqq {}}
\def\eq{Eq.\,}
\def\eqs{Eqs.\,}
\def\beq{\begin{equation}}
\def\eeq{\end{equation}}
\def\bear{\begin{eqnarray}}
\def\bearr{\bear \lal}
\def\ear{\end{eqnarray}}
\def\earn{\nonumber \ear}
\def\nn{\nonumber\\ {}}

\def\nnn{\nonumber\\ \lal }
\def\nnnv{\nonumber\\[5pt] \lal }
\def\yy{\\[5pt] {}}

\def\eql{\al =\al}

\def\tst{\textstyle}

\def\fract#1#2{{\tst\frac{#1}{#2}}}

\def\half{{\fract{1}{2}}}

\def\e{{\,\rm e}}
\def\d{\partial}

\def\flun{\mbox{${\rm cm}^{-2}{\rm s}^{-1}$}\ }

\def\RE{R_{\oplus}}
\def\Horb{H_{\rm orb}}

\def\gsim{\mathrel{
    \raisebox{3pt}{$\mathop{>}\limits_{\displaystyle \sim}$}
		  }}
\def\lsim{\mathrel{
   \raisebox{3pt}{$\mathop{<}\limits_{\displaystyle \sim}$}
		  }}
\def\z{\mbox{$\phantom{0}$}}

\bls{1.03}
\heads
{A.D. Alexeev, K.A. Bronnikov, N.I. Kolosnitsyn,
      M.Yu. Konstantinov, V.N. Melnikov and A.J. Sanders}
      {Measurement of the gravitational constant in space (Project
      SEE)}

\begin{document}
\twocolumn[
\thispagestyle{empty}

\rightline{\bf gr-qc/0104066}

\bigskip

\Title{\bf
Measurement of the gravitational constant $G$ in space (Project SEE): \\
sensitivity to orbital parameters and space charge effect}

\Authors{A.D. Alexeev\foom 1, K.A. Bronnikov\foom 2, N.I. Kolosnitsyn\foom 3,
	    M.Yu. Konstantinov\foom 4, \\ V.N. Melnikov\foom 5}
	    {and A.J. Sanders\foom 6}
{Russian Gravitational Society, 3-1 M. Ulyanovoy St., Moscow 117313, Russia}
{Dept. of Physics and Astronomy, University of Tennessee, Knoxville, TN
		37996-1200, USA}

\medskip

\Abstract
{We describe some new estimates concerning the recently proposed SEE
(Satellite Energy Exchange) experiment for measuring the gravitational
interaction parameters in space. The experiment entails precision tracking
of the relative motion of two test bodies (a heavy ``Shepherd'', and a light
``Particle'') on board a drag-free space capsule. The new estimates include
(i) the sensitivity of Particle trajectories and $G$ measurement to the
Shepherd quadrupole moment uncertainties; (ii) the measurement errors of
$G$ and the strength of a putative Yukawa-type force whose range parameter
$\lambda $ may be either of the order of a few meters or close to the Earth
radius; (iii) a possible effect of the Van Allen radiation belts on the SEE
experiment due to test body electric charging. The main conclusions are
that (i) the SEE concept may allow one to measure $G$ with an uncertainty
smaller than $10^{-7}$ and a progress up to 2 orders of magnitude is
possible in the assessment of the hypothetic Yukawa forces and (ii) van Allen
charging of test bodies is a problem of importance but it may be solved by
the existing methods. }

\medskip
] 

\email 1 {ada@mics.msu.su}
\email 2 {kb@rgs.mccme.ru}
\email 3 {nikkols@orc.ru}
\email 4 {konstmyu@edward.netclub.ru}
\email 5 {melnikov@rgs.phys.msu.su\\
	Temporary address: CINVESTAV, Apartado Postal
	14-740, Mexico 07360, D.F., Mexico;
	e-mail: melnikov@fis.cinvestav.mx}
\email 6 {asanders@utkux.utcc.utk.edu}

\section{Introduction}

The SEE (Satellite Energy Exchange) concept of a space-based gravitational
experiment was suggested in the early 90s \cite{SD92} and was aimed at
precisely measuring the gravitational interaction parameters: the
gravitational constant $G$, possible violations of the equivalence principle
measured by the E\"otv\"os parameter $\eta $, time variations of $G$, and
hypothetical non-Newtonian gravitational forces (parametrized by the Yukawa
strength $\alpha $ and range $\lambda $). Such tests are intended to
overcome the limitations of the current methods of ground-based
experimentation and observation of astronomical phenomena and to fill gaps
left by them. The significance of new measurements is quite evident since
nearly all modified theories of gravity and unified theories predict some
violations of the Equivalence Principle (EP), either by deviations from the
Newtonian law (inverse-square-law, ISL) or by composition-dependent (CD)
gravity accelerations, due to the appearance of new possible massive
particles (partners); time variations of $G$ ($G$-dot) are also generally
predicted \cite{dSMP, Mel94}.

Since gravitational forces are so very small, precision-measurement
techniques have been at the core of terrestrial gravity research for two
centuries. However, evidence is increasingly accumulating which indicates
that terrestrial methods have plateaued in accuracy and are unlikely to
achieve significant accuracy gains in the future. Ref.\,\cite{Gi97}
describes the contemporary situation relative to the measurements of $G$
and $G$-dot.

There is considerable evidence that the uncertainty in $G$ has plateaued at
about 100 ppm \cite{FiArm,Nolting,Meyer,Karag}. Most of the stated
errors in the recent experiments are of the order 100 ppm. Moreover, the
scatter (1-sigma) about the mean is about 140 ppm.
This conclusion does not include the recent paper by Gundlach and
Merkowitz \cite{Gund} who report an error of about 14 ppm in $G$
measurement using a dynamically driven torsion balance. The reported value
of $G$ agrees with the previous measurements within their uncertainties.
This work is unique for the moment and needs confirmation by other
experimental groups using alternative methods.

It might seem that the problems of terrestrial apparatus must inexorably
yield to new technologies --- that the promise of ever increasing
sensitivities would also lead to ever improving accuracy. However, this may
not be true, since it is the intrinsically weak nature of the force
and the resulting systematic errors which arise in its isolation and
measurement that limit the ultimate attainable accuracy in terrestrial
experiments \cite{Gi97}.

The EP may be tested by searching for either violations of the
inverse-square law (ISL) or composition-dependent (CD) effects in
gravitational free fall.

In the watershed year of 1986, Fischbach et al. startled the physics
community by showing that E\"otv\"os's famous turn-of-the-century
experiment is much less decisive as a null result than was generally
believed \cite{Fis86, 12y}.  Prior to this time, experiments by Dicke
\cite{Roll64} and Braginsky \cite{Brag71} had demonstrated the universality
of free fall (UFF) to very high accuracy with respect to several metals
falling in the gravitational field of the Sun (the E\"otv\"os parameter
$\eta$ was ultimately found to be smaller than $ 10^{-12}$). The
interpretation of these results at the time was that they validated UFF.

Since 1986 it has become customary to parametrize possible apparent EP
violations as if due to a Yukawa particle with a Compton wavelength $\lambda
$. This approach unites both ISL and CD effects very naturally, while the
parameter values in the Yukawa potential suggest which experimental
conditions are required to detect the new interaction.

Following the conjecture of Fischbach et al., ISL and CD tests were
undertaken by many investigators. Although a number of anomalies were
initially reported, nearly all of these were eventually explained in terms
of overlooked systematic errors or extreme sensitivity to models, while
most investigators obtained null results. However, a positive result for a
deviation from the Newtonian law (ISL) was obtained (and interpreted in
terms of a Yukawa-type potential) in the range of 20 to 500 m by Achilli
and colleagues \cite{Ach}; this needs to be verified in other independent
experiments.

For reviews of terrestrial searches for non-Newtonian gravity, see \cite
{Adel94, FiG, Franklin}.

The idea of the SEE method is to study the relative motion of two bodies on
board a drag-free Earth satellite using horseshoe-type trajectories,
previously well-known in planetary satellite astronomy: if the lighter body
(the ``Particle'') is moving along a lower orbit than the heavier one (the
``Shepherd'') and approaching from behind, then the Particle almost
overtakes the Shepherd, but it gains energy due to their gravitational
interaction, passes therefore to a higher orbit and begins to lag behind.
The interaction phase can be studied within a drag-free capsule (a cylinder
up to 20 m long, about 1 m in diameter) where the Particle can loiter as
long as $10^5$ seconds. It was claimed that the SEE method exceeded in
accuracy all other suggestions, at least with respect to $G$ and $\alpha $
for $\lambda $ of the order of meters. Some design features were considered,
making it possible to reduce various sources of error to a negligible level.
It was concluded, in particular, that the most favorable orbits
are the sun-synchronous, continuous sunlight orbits situated at altitudes
between 1390 and 3330 km \cite{SD92}.

Since the origination of the SEE concept, the development has focused on
critical analyses of orbital parameters and satellite performance and
the assessment of critical hardware requirements.  All indications from
this work are that the SEE concept is feasible and practicable
\cite{iztech}.

At the present stage one can assert that,
although space is a challenging environment for research, the
inherent quietness of space can be exploited to make very accurate
determinations of $G$ and other gravitational parameters, providing that
care is taken to understand the many physical phenomena in space which have
the potential to vitiate accuracy. A distinctive feature of a SEE mission is
its capability to perform such determinations simultaneously on multiple
parameters, making it one of the most promising proposals.

To be more specific, let us enumerate the suggested SEE tests and
measurements and show their expected accuracy as currently estimated:

\medskip\noi\
\begin{tabular}{ll}
{\it Test/measurement} \cm&    {\it Expected accuracy}
										\\[8pt]
EP/ISL at a few metres  &       $2\ten{-7}$               \yy
EP/CD at a few metres   &       $<10^{-7}\ (\alpha < 10^{-4})$ \yy
EP/ISL at $\sim R_{\oplus}$     &       $<10^{-10}$          \yy
$G$                     &       $3.3\ten{-7}$                 \yy
$\dot G/G$              &       $<10^{-13}$ in one year
\end{tabular}
\medskip

The last estimate is only tentative; the subject is under study.

This paper presents a description of the main ideas of the SEE experiment
and some new evaluations concerning the opportunities of the SEE concept
and its yet-unresolved difficulties. In \sect 2, on the basis of computer
simulations of Particle trajectories, we describe some general properties of
Particle trajectories in short (Particle-Shepherd separation from 2 to 5
meters) capsules, their sensitivity to the value of the parameters of the
gravitational interaction, and we estimate the requirements to the Shepherd
quadrupole moment uncertainty. \sect 3 shows the results of simulations of
the measurement procedure itself, which enables us to estimate the possible
measurement accuracy with respect to $G$ and $\alpha $ for $\lambda $ of the
order of either meters or the Earth's radius. \sect 4 discusses a spurious
effect of test body electric charging when the satellite orbit passes
through the Van Allen radiation belts, rich in high-energy protons. \sect 5
is a conclusion.

In what follows, the term ``orbit" applies to satellite (or Shepherd) motion
around the Earth, while the words ``trajectory" or ``path" apply to Particle
motion with respect to the Shepherd inside the capsule.

\section{Simulations of Particle trajectories}

In the previous studies of the SEE project it was assumed that the capsule
was about 20 m long and the initial Shepherd-Particle separation $x_0$ along
the capsule axis was as great as 18 m; some estimations were also made for $
5 $ m $\leq x_0\leq 10$ m. The Shepherd mass was taken to be $M=500$ kg and
the Particle mass $m=0.1$ kg. The present study retains these values,
except the cases noted below.

\onecol
In this section we will describe some characteristic features of Particle
trajectories in a short capsules (Particle-Shepherd separation $2$ m $\leq
x_0\leq 5$ m) with the Shepherd mass reduced to $M=200$ kg. Our goal is
to determine the properties of trajectories in the case
considered and to determine the sensitivity of trajectories to the
uncertainty of orbit radius, the value of the Newtonian gravitational
constant and of the Shepherd quadrupole moment $J_2$. As in our previous
studies, the capsule diameter is presumed to be 1 m.

The reason for considering the short capsule, the Shepherd with reduced
mass and the quadrupole moment uncertainty is economical and technological
in origin. Namely, it is hard to produce a spherically symmetric
Shepherd to the required accuracy. To avoid the inclusion of $\delta J_2$
in the set of parameters to be determined in the experiment, it is useful
to know which values of $ \delta J_2$ will be negligible, since the growth
of the number of parameters leads to serious problems in data processing.

\subsection{Equations of motion and initial data}

For simplicity, we assume that the relative motion of the test
bodies inside the capsule occurs in the satellite orbital plane. This
simplification is purely technical, since, as was found in our
previous studies, the three-dimensional nature of the Particle motion does
not change the main estimates.

The reduced Lagrangian of the Particle motion in the considered case is
\bearr
\label{lagr}L=\frac M2(\dot R^2+R^2\dot \varphi ^2)+\frac m2\left[ \dot
r^2+r^2(\dot \varphi +\dot \psi )^2\right]
+G\frac{M_{\oplus }m}r+G\frac{Mm}
s\left\{ 1+J_2\left( \frac{r_s}s\right) ^2\!P_2(\cos \theta )\right\} \left(
1+\alpha e^{-s/\lambda }\right)
\ear
where $(R,\varphi )$ are the Earth-centered polar coordinates of the
Shepherd in the orbital plane; $r=\sqrt{(R{+}y)^2{+}x^2}$ and $\psi $ are
the Earth-centered polar coordinates of the Particle; $x$ and $y$ are the
Shepherd-centered Particle coordinates, where $x$ is the ``horizontal'' one,
i.e., along the orbit and simultaneously along the capsule and $y$ is the
``vertical'' one, along the Earth-Shepherd radius vector; $s=\sqrt{x^2+y^2}$
is the Particle-Shepherd separation; $M_{\oplus }$, $M$ and $m$ are the
Earth, Shepherd and Particle masses, respectively; $J_2$ is the quadrupole
moment of the Shepherd, $r_s$ is its radius and $P_2$ is the Legendre
polynomial
\[
	P_2(\cos \theta )=\frac{3\cos ^2\theta -1}2,
\]
where $\theta $ is the angle between the line connecting the centres of
test bodies and the Shepherd equatorial plane. It is easy to see that if the
Shepherd symmetry axis is in its orbital plane, then $\theta =\theta
_0=-\arctan (y/x)+\varphi $. If the symmetry axis of the Shepherd is
orthogonal to its orbital plane, then $\theta =0$. In general, if $\chi $
is the angle between the Shepherd symmetry axis and its orbital plane, then
$\theta =\theta _0\cos \chi $. Hence the influence of $J_2$ on the Particle
motion is minimum if the Shepherd symmetry axis lies in its orbital plane
and is maximum if they are mutually orthogonal.

For simplicity (and taking into account the corresponding estimate) we
neglect the influence of the Particle on the Shepherd, so the Shepherd
trajectory is considered as given. Then, varying the above Lagrangian
with respect to $x$ and $y$, taking into account that $M\gg m$ and $R\gg s$,
we arrive at the following equations of Particle motion with respect to the
Shepherd:
\bearr
\frac{d^2x}{dt^2}=2\dot y\dot \varphi +x\left\{ \dot \varphi ^2-\frac{
GM_{\oplus }}{r^3}\right\} -\frac{2\dot R\dot \varphi y}R-\frac{G\overline{M}
}{s^3}x\left\{ 1+J_2\left( \frac{r_s}s\right) ^2\!P_2(\cos \theta )\right\}
\nnn \cm\cm
-\alpha x\frac{G\overline{M}}{s^2}\left\{ 1+J_2\left( \frac{r_s}s\right)
^2\!P_2(\cos \theta )\right\} \!\left( \frac 1s+\frac 1\lambda \right)
\e^{-s/\lambda }
\nnn \hspace{5cm}                                            \label{eqnx}
+\frac{G\overline{M}r_0^2}{2s^5}J_2\left( 1+\alpha \e^{-s/\lambda
}\right) \times \left[ x(1+3\cos 2\theta )+3y\sin 2\theta \cos \chi \right];
\\ \lal
\frac{d^2y}{dt^2}=-2\dot x\dot \varphi +(R+y)\left\{ \dot \varphi ^2-\frac{
GM_{\oplus }}{r^3}\right\} +\frac{2\dot R\dot \varphi x}R-\frac{G\overline{M}
}{s^3}y\left\{ 1+J_2\left( \frac{r_s}s\right) ^2\!P_2(\cos \theta )\right\}
\nnn \cm\cm
- \alpha y\frac{G\overline{M}}{s^2}\left( \frac 1s+\frac 1\lambda \right)
\!\left\{ 1+J_2\left( \frac{r_s}s\right) ^2\!P_2(\cos \theta )\right\}
\e^{-s/\lambda }
\nnn \hspace{5cm}                                            \label{eqny}
+\frac{G\overline{M}r_0^2}{s^5}J_2\left( 1+\alpha \e^{-s/\lambda
}\right) \times \left[ 3x\sin 2\theta \cos \chi +y(1-3\cos \theta )\right]
\ear
where $\overline{M}=M+m$.

\twocol
Two kinds of initial conditions for \eqs (\ref{eqnx}) and (\ref{eqny}) were
used during the simulations. First, we used the so-called ``standard''
initial conditions, taking the Particle velocity components $\dot x(0)$ and $
\dot y(0)$ corresponding to its unperturbed (i.e., without the $M-m$
interaction) orbital motion distinguished from the Shepherd's orbit only by
its radius (for circular orbits) or semimajor axis (for elliptic orbits).
Assuming that the Particle motion begins right at the moment when the
Shepherd passes its perigee, these conditions have the form
\bear                                               \label{initcond}
		x(0) \eql x_0,\qquad y(0)=y_0,
\nn
	    \dot x(0)\eql \frac{\omega e' y_0}{2(1-e)^2},
\qquad
	    \dot y(0) = -\frac{\omega ex_0}{e'(1-e)}
\ear
where $\omega ^2=GM_{\oplus }/R_0^3$, $R_0$ is the Shepherd orbital radius
(at the perigee), $e$ is the orbital eccentricity and $e'=\sqrt{1-e^2}$.

For clearness, the relations (\ref{initcond}) are written in the linear
approximation in the variables $x$ and $y$. Higher-order approximations were
used in the simulation process as well.

The second kind of initial conditions corresponds to small variations of
initial velocities with respect to their ``standard'' values.

The set of equations (\ref{eqnx})--(\ref{eqny}) was solved numerically using
the software developed previously \cite{iztech} to analyze the SEE project.

On the basis of numerical solutions of \eqs (\ref{eqnx}) and (\ref{eqny}),
we considered two types of Particle trajectories, corresponding to
different choices of the initial data: (i) approximately U-shaped
ones and (ii) cycloidal ones, containing loops (see more details on
the trajectories in \cite{SD92,iztech}), for orbital altitudes $H_{{\rm
orb}}=500$, 1500 and 3000 km.

\subsection{General characteristics of trajectories in a short capsule}

Reduction of the Shepherd mass leads to the reduction of distance between
libration points $L_1$ and $L_2$. As a result, the region where the
horseshoe orbits exist is reduced as well, and the turning points for the
horseshoe orbits become closer to the Shepherd. Thus, for a Shepherd mass
of $ M=200$ kg, the positions of the turning points of the horseshoe orbits
starting, for instance, at $x_0=18$ m and $0.1$ m$\leq |y_0|\leq 0.3$ m are
placed in the region $x\leq 5$ m for all orbits whose altitude $H$ is in
interval $500$ km $\leq H\leq 3000$ km. The usage of short trajectories,
which start in the region $x_0\leq 5$ m, and especially extremely short
trajectories with $x_0=2$ m, leads to additional limitations on the initial
conditions.

Numerical simulation shows that to avoid Particle collisions with the
Shepherd and the walls of the capsule for ''smooth'' trajectories, which
correspond to the standard initial conditions, trajectories with
$|y_0|\leq 0.2$ m may be used for orbit altitudes $\Horb \leq 1500$
km while for orbit altitude $\Horb =3000$ km and $x_0\geq 3$ m
trajectories with $ |y_0|=25$ cm may be used also.

The usage of cycloidal trajectories, which appear when the absolute values
of the Particle's initial velocity exceed its standard value, make it
possible to avoid this limitation.

\subsection{Trajectory sensitivity with respect to the Newtonian
		  gravitational constant}

The influence of the Newtonian gravitational constant $G$ on the Particle
motion in the case considered may be investigated by considering how
a small perturbation $\delta G$ from the ``standard'' value of
$G$ changes the Particle trajectories. Such a perturbation is characterized
by the displacement of a perturbed trajectory with respect to an unperturbed
one, that is,
\[
	\delta \vec r =\vec r_\delta (t)-\vec r_0 (t)
\]
where $\vec r_\delta$ and $\vec r_0$ denote the Particle radius
vector for perturbed and unperturbed motion, respectively.
For rather long trajectories, instead of the full displacement $\delta
\vec{r}$, the displacement along the $x$ axis ($\delta x$) may
be considered, because a numerical simulation shows that a displacement
along the $y$ axis is one order of magnitude smaller than a displacement
along the $x$-axis.

U-shaped Particle trajectories were considered for circular Shepherd
orbits in the following range of parameters and initial conditions:
orbit altitudes $\Horb= 500$, $1500$ and $3000$ km; the initial position
changes in the range $2$ m $\leq x_0\leq 5$ m, $-20$ cm $\leq y_0\leq -5$
cm. For comparison with the case where $M=500$ kg,  trajectories with
$x_0=18$ m were considered also. It was found that the perturbation $\delta
G/G=10^{-6}$ leads to displacement of smooth trajectories in the range from
$0.634\cdot 10^{-6}$ m to $3.34\cdot 10^{-6}$ m with a minimum displacement
achieved for $y_0=0.05$ m, $x_0=2$ m and $\Horb =3000$ km while a maximal
displacement is achieved for $y_0=0.05$ m, $x_0=5$ m and $\Horb = 500$ km.

The use of cycloidal trajectories increases the influence of $\delta G$
on the Particle motion.

\subsection
{Trajectory sensitivity to the orbit radius uncertainty}

This sensitivity is characterized by the displacement $\delta x$, induced by
a small perturbation (or uncertainty) $\delta h$ of the orbital altitude
\footnote{As above, we consider the displacement $\delta x$ instead of
$\delta r$ because $\delta x$ provides the main contribution to $\delta r$,
while $\delta y$ is much smaller.} $\Horb$.

It was found that for $\delta h=1$ cm, the orbital altitude $\Horb =1500$ km
and $ |y_0|=20$ cm, the maximum value of $\delta x$ increases from
$3.222\cdot 10^{-8}$ m at $ x_0=2$ m to $9.387\cdot 10^{-8}$ m at $x_0=5$ m.
For $\Horb =500$ km these values must be multiplied by a factor of 2
(approximately) and for $\Horb =3000$ km by a factor of 1/2.

It was also found that the dependence $\delta x(\delta h)$ is, to a good
accuracy, linear: in particular, for $\delta h=1$ m,
the orbital altitude $\Horb =1500$ km, $x_0=5$ m and $|y_0|=20$ cm, $\delta
x=9.387\cdot 10^{-6}$ m as expected.

The use of cycloidal trajectories reduces the dependence of trajectories
on the orbit altitude: for instance, for $\Horb =1500$ km,
$ x_0=5$ m and $|y_0|=20$ cm, the uncertainty $\delta h=1$ cm leads to
$\delta x=1.455\cdot 10^{-8}$ m.

It is necessary to point out that the influence of the orbital altitude on
the Particle trajectory for the reduced Shepherd mass ($M=200$ kg) is
greater than in the case $M=500$ kg. For instance, in the case
$M=500$ kg, $\Horb =1500$ km, $\delta h=1$ m, $x_0=5$ m and $|y_0|=20$ cm,
the trajectory displacement becomes $\delta x=4.6593\cdot 10^{-8}$ m.

These estimates lead to certain restrictions on the reasonable precision of
Particle trajectory measurements.

\subsection{Trajectory sensitivity to the Shepherd's
quadrupole moment uncertainty}

A maximum effect of the Shepherd quadrupole moment $J_2$ on the
Particle motion is realized in the case when the Shepherd axis is
orthogonal to its orbital plane. The influence of $\delta J_2$ on the
accuracy of $G$ measurement may be estimated as follows. Let some value of
$\delta J_2 $ produce the trajectory displacement $|\delta \vec r|\leq
\delta l_j$ while the variation $\delta G_0$ of $G$ with the same initial
conditions gives the trajectory displacement $|\delta \vec r|\leq \delta
l_G$. Then, keeping in mind the linear dependence of trajectory
displacements on $\delta J_2$ and $\delta G$, the accuracy of $G$
measurement under the uncertainty $ \delta J_2$ may be estimated as
\[
   \frac{\delta G} G\leq
    \biggl( 1 + \frac{\delta l_j}{\delta l_G}\biggr)\frac{\delta G_0}G.
\]

Using this inequality and the results of trajectory simulations, we obtain
the following estimates for U-shaped Particle trajectories in circular
orbits with $H_{{\rm orb}}=1500$ km:

\begin{table}
{\bf Table 1. } Estimates of $\delta G/G$ in ppm for $\delta J_2=10^{-4}$,
when the symmetry axis of the Shepherd is orthogonal ($\chi =\pi /2$) to its
orbital plane. The second line shows $x_0$.

\begin{center}
\begin{tabular}{|r|r|r|r|r|}
\hline
$y_0,$ &
      \multicolumn{4}{|c|}{$x_0$} \\
\cline{2-5}
 cm  & 2 m & 3 m & 4 m &
5 m \\
\hline
	-5 & $1.568$ & $5.13$  &  $2.41$ & $0.138$
\\
\hline
	-10 & $3.181$ & $1.41$  &  $0.876$ & $0.649$  \\
\hline
	-15
& $0.857$ & $6.08$  &  $5.057$ & $4.486$ \\
\hline
              -20 & $
60.28$ & $44.31$  &  $31.91$ & $28.04$ \\
\hline
\end{tabular}
\end{center}
\end{table}

The uncertainties $\delta J_2 \lsim 10^{-5}$
do not create a substantial error in $G$ for most of the trajectories. Only
trajectories with $|y_0|=0.2$ m require $\delta J_2 \lsim 10^{-6}$
because of the growth of the sinusoidal component for these trajectories.

\section{Simulations of experimental procedures}

This section describes the results of numerical simulations of the whole
measurement procedures aimed at obtaining the sought-after gravitational
interaction parameters. These simulations assumed a Shepherd mass of $M=500$
kg, a circular orbit with $H_{{\rm orb}}=1500$ km with a spherical
gravitational potential for the Earth, and a Particle mass of $100$ g. Where
relevant, it is assumed that both the Shepherd and the Particle are made of
tungsten. Identical compositions for them are assumed for simplicity since
this work is performed for estimation purposes only.

\subsection{Simulations of an experiment for measuring $G$}

The constant $G$ is determined from the best fitting condition between the
``theoretical'' Particle trajectories ($\vec r{}^{{\rm \ th}}(t_i)=\vec
r{}_i^{{\rm \ th}}$), calculated by Eqs. (2) and (3) for $J_2=0$ and the
``empirical'' ($\vec r_i$) Particle trajectories near the Shepherd. The
fitting quality is evaluated by minimizing a functional that characterizes
a ``distance'' between the trajectories. We have considered the following
functionals for such ``distances'':
\bearr
     S = \sum_{i=1}^N \biggl [(x_i-x{}_i^{{\rm th}})^2
		+ (y_i-y{}_i^{{\rm th }})^2\biggr], 	 	\label{F1}
\\ \lal
	S_x= \sum_{i=1}^N (x_i-x{}_i^{{\rm th} })^2,
\qquad
	S_y= \sum_{i=1}^N (y_i-y{}_i^{{\rm th}})^2, 	 	\label{F2}
\ear

The theoretical trajectory depends on the gravitational constant $G$, on the
initial coordinates $x_0,y_0$ and on the initial velocities $v_{x0},v_{y0}$.
To estimate $G$, one chooses the value for which a ``distance'' functional
in the space of the five variables ($G,x_0,y_0,v_{x0},v_{y0}$) reaches its
minimum.

\begin{figure}
\centerline{\epsfxsize=83mm \epsfbox{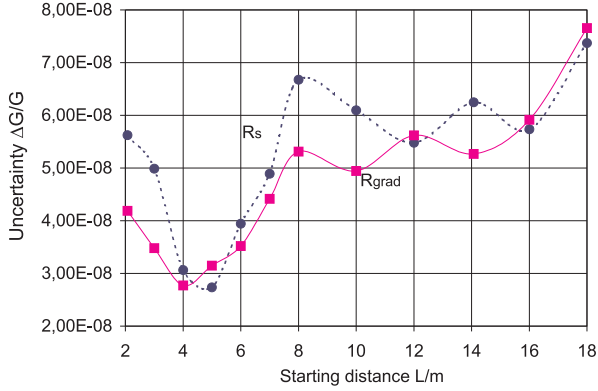}}
\caption{\protect\small
Uncertainties $\delta G$ estimated by the gradient descent
($R_{\rm grad}$) and consecutive descent ($R_{\rm s}$) methods.}
\medskip \hrule
\end{figure}

We carried out a numerical simulation of the SEE experiment and estimated $
\delta G$ for a given coordinate measurement error ($\sigma=1\ten{-6}$ m).
As ``empirical'' trajectories, we took computed trajectories, with specified
values of the above five variables, where Gaussian noise was introduced
from a random number generator. Independent ``empirical'' trajectories were
created by non-intersecting random number sequences. The functional was
minimized using the gradient descent method and the consecutive descent
method. The starting value of the ``vertical'' (along the Earth's radius)
coordinate, $y_0$, was taken to be 0.25 m, while the horizontal one, $x_0$,
varied between 2 and 18 m. Fig\,1
shows the dependence of the errors $\delta G/G = R_{{\rm grad}}$, obtained
by the gradient descent method and $\delta G/G = R_{{\rm s}}$, obtained by
the consecutive descent method. All the errors are estimated by confidence
intervals corresponding to a confidence of 0.95. The mean values of these
errors are as follows: \bear
R_{{\rm grad}} = 4.69\ten{-8},\cm
R_{{\rm s}} \eql 5.24\ten{-8}. \earn
Thus the errors estimated by the gradient and consecutive descent methods
are close to each other and are about an order of magnitude smaller than the
error from one-trajectory data. It was found that the simulation
results depend strongly on the random number generator, so that ordinary
generators are not perfect: the generated random number sets do not obey
the Gaussian law.

The use of truncated functionals like (2) has shown that a functional
incorporating the more informative ``horizontal'' coordinate $x$ leads to
estimates close to those obtained from the total functional, whereas the use
of $y$ alone substantially decreases the sensitivity. Therefore in practice,
to determine $G$, it is sufficient to measure only one of the two
coordinates, viz. $x$.

Since the ``empirical'' trajectory is built on the basis of a computed one,
with a known value of the gravitational constant $G_0$, it appears possible
to estimate a possible systematic error inherent in the data processing
method. The latter has turned out to be in most cases much smaller than the
random error. This result shows the correctness of the methods used.

As is evident from the results, the best accuracy is achieved at values of $
x_0$ ($\approx $ the capsule size) about 4--5 meters.

\begin{figure}
\centerline{\epsfxsize=83mm \epsfbox{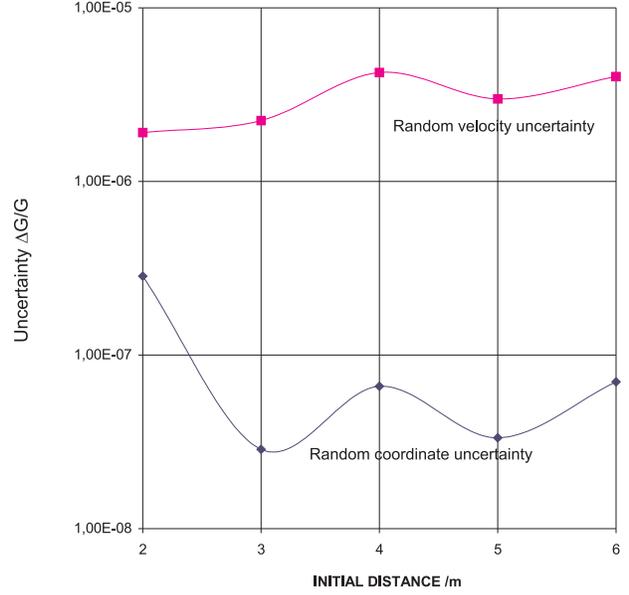}}
\caption{\protect\small
Random ``velocity" and ``coordinate" uncertainties.}
\medskip \hrule
\end{figure}

\subsection{Simulation of measurement procedures for estimation
		of $G$ using velocity data}

When one can precisely measure the Particle velocity, the
gravitational constant can be estimated from the velocity data alone.
A computer program was created for simulating an experiment
determining the gravitational constant $G$ by velocity data. These
simulations assumed a Shepherd mass of $500$ kg, a circular orbit with
$\Horb=1500$ km (spherical Earth's potential), a Particle mass of
$100$ g, and a velocity error $\delta v_x=1\cdot
10^{-8}$ m/s.  The constant $G$ is estimated from the best
fitting condition between the theoretical Particle velocity $\left(
\vec v^{\rm th}(t_i)\equiv \vec v_i^{\rm th}\right) $ and the empirical one.
The fitting quality is evaluated by minimizing a functional characterizing a
distance between velocity trajectories. We used the following functional:

\[
	V=\sum_{i=1}^N\left[ \left( v_{xi}-v_{xi}^{\rm th}\right) ^2+\left(
		v_{yi}-v_{yi}^{\rm th}\right) ^2\right] .
\]

The theoretical velocity depends on the gravitational constant $G$, on the
initial coordinates $x_0,y_0$ and the initial velocities $v_{x0},v_{y0}$. To
estimate $G$, one chooses the value for which the velocity distance
functional $V$  reaches its minimum in the space of the five variable
($G,x_0,y_0,v_{x0},v_{y0}$)

As empirical trajectories, we took computed trajectories with specified
values of the above five variables, where Gaussian noise was introduced
from a random number generator. Independent empirical trajectories were
created by non-intersecting random number sequences. The functional $V$ was
minimized using the gradient descent method. The starting value of the $y_0$
coordinate was taken to be $0.25$ m, while the horizontal one, $x_0$, varied
between $2$ and $6$ m. The value of $G$ was estimated from 11
trajectories at a significance level of 0.95. The appropriate errors were
estimated at confidence intervals and then related to $G$
as shown by the upper curve in Fig. 2.  The relative error obtained using
the coordinate functional for a coordinate measurement error of $\sigma
=1\cdot 10^{-6}$ m is also shown in the same figure. Computer-based
simulation also allows one to estimate systematic errors.  We see that the
accuracy of estimation of $G$ for a given set of measurement errors is the
best for the coordinate functional.

\subsection{Equations of motion with Yukawa terms}

We will present the Particle equations of motion in the
relevant approximation, including the contributions from hypothetical Yukawa
forces, taking into account the finite size of the Yukawa field sources.

Let the interaction potential for two elementary masses $m_1$ and $m_2$ be
described by the potential
\beq                                                         \label{Yuk1}
	dV^{{\rm Yu}}= \frac{G\,dm_1dm_2}r \alpha \e^{-r/\lambda }
\eeq
where $r$ is the masses' separation, $\alpha $ and $\lambda $ are the
strength parameter and the range of the Yukawa forces. Then for two massive
bodies with the radii $R_1$ and $R_2$ after integration over their volumes
we obtain \cite{ZaKol}
\beq
V^{{\rm Yu}}= \frac{G\,m_1m_2\beta _1\beta _2}r \label{Yuk2} \alpha \e
^{-r/\lambda }
\eeq
where
\beq                                                         \label{Yuk3}
\beta _i= 3 \biggl(\frac \lambda {R_i}\biggr)^3\biggl[\frac{R_i}
\lambda \cosh \frac{R_i}\lambda - \sinh \frac{R_i}\lambda \biggr].
\eeq
When $R_i/\lambda \ll 1$, we have $\beta _i\approx 1$. This may be the case
when we consider the interaction between the Shepherd and the Particle at a
distance of the order of a few meters. The radii of the Shepherd and the
Particle are small: $R_1\approx 18$ cm for the Shepherd and $R_2\approx 1.1$
cm for the Particle. If the range $\lambda $ is of the order of the Earth
radius, $\lambda \approx R_{\oplus }$, we have $\beta _{\oplus }=1.10$ and $
\beta _{1,2}=1$ where the indices 1 and 2 label the Shepherd and the
Particle, respectively.

Equations of motion are obtained under the following assumptions. There
are two Yukawa interactions with the parameters $\lambda_0$ and $\alpha_0$
referring to the Earth-Shepherd and Earth-Particle interactions which are
the same (due to the assumed identical composition for the Shepherd and the
Particle), while $\lambda$ and $\alpha$ determine the Shepherd-Particle
interaction. The equations of motion in the frame of reference of the
Shepherd, with the same notations for $x$, $y$ and $s$ as used previously,
are
\bearr
	\ddot{x} +2\omega^2 \dot{y} + G(m_1+m_2) \frac{x}{s^3}
		- 3\omega^2 \frac{xy}{s}
\nnn \cm
	+ G(m_1+m_2) \frac{x}{s^3} \alpha \biggl(1+ \frac{s}{\lambda}\biggr)
	\e^{-s/\lambda} =0;
\nnnv
   \ddot{y} -2 \omega\dot{x} - 3\omega^2 y + G(m_1+m_2)\frac{y}{s^3}
    + \frac{ 3\omega^2}{r_{01}} \biggl(y^2 -\frac{x^2}{2}\biggr)
\nnn
    +G(m_1+m_2)\frac{y}{s^3} \alpha \biggl(1+\frac{s}{\lambda}\biggr)\e
        ^{-s/\lambda}
\nnn
	\inch - \omega^2 \beta_0 \alpha_0 \e^{-r_{01}/\lambda_0}y =0
                                                         \label {Yuk4}
\ear
where $\omega$ is the orbital frequency:
\beq                                                          \label{omega}
    \omega^2 = \frac{GM_{\oplus}}{r_{01}^3}
    \biggl[ 1+\beta_0 \alpha_0   \biggl(1+ \frac{r_{01}}{\lambda_0}\biggr)
    					\e^{-r_{01}/\lambda_0}\biggr].
\eeq
We have neglected the terms quadratic in $s/r_{01}$ times $\alpha$ or $
\alpha_0$ due to their manifestly small contributions.

If we set $\alpha_0=0$ in \eqs (\ref{Yuk4}), we obtain the equations used to
describe only the Shepherd-Particle Yukawa interaction. Notice that the
Yukawa terms are roughly proportional to the gradients of the corresponding
Newtonian accelerations, namely, $Gm_1/s^3$ for the Shepherd-Particle
interaction and $GM_{\oplus}/r_{01}^3 \approx \omega^2$ for (say) the
Earth-Shepherd interaction. In our case these quantities are estimated as
\bearr
    \frac{Gm_1}{s^3} \approx 2.7\ten{-10}\ {\rm s}^{-2} \cm
		{\rm for}\ \ s=5\ {\rm m},
\nnn   \inch
		{\rm and}\quad \omega^2 \approx 8.16\ten{-7} \ {\rm s}^{-2}.
\ear
Thus, given the same strength parameter, the Earth's Yukawa force is three
orders of magnitude greater than that between the Shepherd and the
Particle.  Therefore, one might expect some significant progress in an ISL
test for $ \lambda$ of the order of the Earth's radius.

\eqs (\ref{Yuk4}) were used to simulate the measurement procedures.

\subsection{Sensitivity to Yukawa forces with $\lambda \sim 1$ m}

In an experiment for finding the Yukawa interaction between the Shepherd and
the Particle using the potential (\ref{Yuk2}) with $\beta _{1,2}=1$, one
computes two theoretical trajectories: the first ignoring the Yukawa forces
($x^0(t_i),\ y^0(t_i)$) and the second taking them into account
$\Bigl(x^\alpha (t_i),\ y^\alpha (t_i)\Bigr)$. These two computed curves
are compared with the empirical trajectory using the functional $S_k$
($k=0,\alpha $) according to (\ref{F1}) which may be considered as a
dispersion characterizing a scatter of the ``empirical'' coordinates with
respect to the fitting trajectory. This is true when the theoretical model
is adequate to the real situation. In the case $k=\alpha $ the functional
$S_k=s_\alpha $ has a $\chi ^2$ distribution with $n_2=2N-1$ degrees of
freedom. With $k=0$ the parameter $\alpha $ is absent, therefore $S_0$ is
distributed according to the $\chi ^2$ law with $N_1=2N$ degrees of
freedom. Then their ratio $ S_0/S_\alpha =F_{n_2,n_1}$ will be distributed
according to the Fischer law \cite{fischlaw}
with $n_2$ and $n_1$ degrees of freedom. If an
experiment shows that, on a given significance level $q$, the relation
(see \cite{12y})
\beq
	S_0/S_\alpha \geq F_{n_1,n_2,q}\label{F5}
\eeq
is valid, one should conclude that the Yukawa force has been detected. An
equality sign shows a minimum detectable force for the given significance
level $q$. We have assumed $q=0.95$. The results of a sensitivity
computation for different values of the space parameter $\lambda $ are
presented in Fig.\thinspace 3 (curve 1). A maximum sensitivity of $\alpha
=2.1\ten{-7}$ has been observed for $\lambda =1.25$ m. This value is 3 to 4
orders of magnitude better than the sensitivity of terrestrial experiments
in the same range.

\begin{figure*}
\centerline{\epsfbox{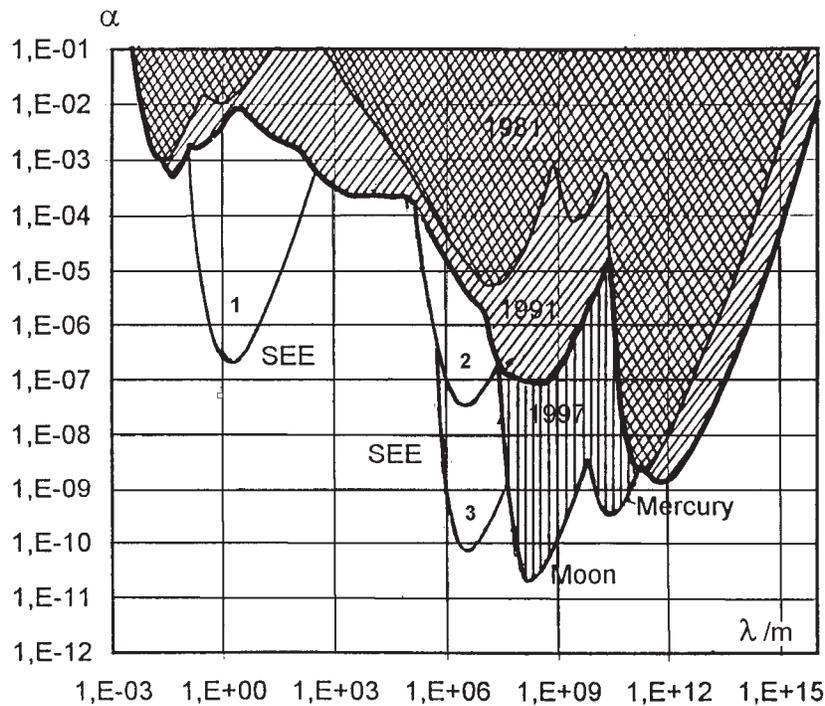}}
\caption{\protect\small
The SEE method sensitivity to Yukawa forces with the range
parameter $\lambda_0$ of the order meters (1) of the Earth's radius $\RE$
using trajectory measurements (2) and orbit precession (3). The limitations
on Yukawa's forces parameters from \protect\cite{12y} (4).}
\medskip \hrule
\end{figure*}

\subsection{Sensitivity to Yukawa forces with $\lambda \sim R_{\oplus}$}

To estimate the parameter $\alpha _0$ in \eqs(\ref{Yuk4}), computer
simulations were carried out using the method as described above for $\alpha
$, based on the Fischer criterion for the significance level 0.95. The range
parameter $\lambda _0$ varied from $(1/32)R_{\oplus }$ to $32R_{\oplus }$.
Two trajectories with the initial Shepherd-Particle separations $x_0$ of 2
and 5 m were calculated. In both cases the impact parameter $y_0$ was chosen
to be 0.25 m. We used \eqs (\ref{Yuk4}) with $\alpha =0$, i.e., excluding
the non-Newtonian interaction between the Shepherd and the Particle. As is
evident from \eqs (\ref{Yuk4}), the Particle trajectory depends on the ratio
$r_{01}/\lambda _0$ in the product $(r_{01}/\lambda _0)\e^{-r_{01}/\lambda
_0}$. This quantity reaches its maximum at $\lambda _0=r_{01}/2$. Our
calculations have confirmed that a maximum sensitivity of the SEE method ($
3.4\ten {-8}$ for $x_0=5$ m) is indeed observed at this value of $\lambda _0$
. This is about an order of magnitude better than the estimates obtained by
other methods. Hopefully this estimate may be further improved by about an
order of magnitude by optimization of the orbital parameters. However, there
is a factor which can, to a certain extent, spoil these results, namely, the
uncertainty in the parameter $\omega $ which, in this calculation, was
assumed to be known precisely.

The simulation results are shown in Fig.\,3 (curve 2) for a
trajectory with an initial Shepherd-Particle separation of 5 metres.
See the next section for further discussion of this figure.

\subsection{Precession of the Shepherd orbit and a test of the Inverse
Square Law at distances of the order of $R_\oplus$}

We have shown previously that the SEE experiment allows one to test the
inverse square law at distances on the order of 1 m ($\alpha_{\min}
\sim 2\cdot 10^{-7}$) and at distances on the order of a half of an orbit
radius ($\alpha_{\min} \sim 3.4\cdot 10^{-8}$) \cite{gracos}. Another test
can be done using spacecraft precession data. As known, in two-body
problems an orbit is closed for only two potentials. They are (1) the
Newtonian potential, $U \sim 1/r$, and (2) $U\sim r^2$. In other cases the
orbit is not closed and a pericenter precession is observed. In particular,
any deviation from the Newtonian law entails a precession of an orbit due
to the Yukawa interaction
\beq
	U'=\frac{Gm_1m_2}r\alpha \e^{-r/\lambda }
\eeq
the Shepherd orbit exhibits a precession. In the
general case the precession magnitude due to a small perturbation,
described by a potential $ \delta U$, is given by (see [1])
\beq
\delta
	\varphi =\frac \d {\d M}\left( \frac{2m'} M\int_0^\pi
			r^2\delta U d\varphi \right)
\eeq
Integration is done over a non-perturbed
trajectory. Here $m$ is the Shepherd's mass, $m'$ is the mass of a central
body (the Earth), $M=mr^2 \dot\varphi$ is an integral of
motion (the angular momentum), and $\delta U=\alpha (Gmm')\e^{-r/\lambda}$.
The non-perturbed trajectory is described by the expressions
\bear
	r \eql \frac p{1+e\cos \varphi },\cm    p=\frac{M^2}{m(Gmm')}
\nn
	e^2 \eql 1+\frac{2EM^2}{m(Gmm')^2}p=a(1-e^2).
\ear

After a standard algebraic computation we obtain
\bearr
  \delta \varphi =\alpha \frac 2e\int_0^\pi \frac{\e^{-r/\lambda }}{(1+e\cos
   \varphi )^2}
\nnn
   \times\left\{ \frac 1\lambda \left[ 2e+(1+e^2)\cos \varphi \right]
      	-(e+\cos \varphi )\right\} d\varphi ,
\ear
where
\beq  \cm
   \frac r\lambda =\frac a\lambda \cdot \frac{1-e^2}{1+e\cos \varphi }.
\eeq
Using \eq (8) and data with the error $\delta \varphi$ for the
SEE Satellite, we calculated the curves $\alpha (\lambda )$, which
determine the border on the $\alpha - \lambda $ plane between two
domains, where the Yukawa interaction (a new long-range force) is forbidden
by experiment and where it is not. The sensitivities to Yukawa interactions
are shown in Fig.\,3 as the curve 4 (see \cite{Fis86}) for the parameter
$\lambda $ in the range from $1\cdot 10^6$ m to $1\cdot 10^{13}$ m. The
curve 3 was calculated for the SEE satellite with an eccentricity $e=0.01$
and the precession error is equal to $\delta \varphi =0.1''/y$.

One can conclude that the inverse square law may be
tested with a sensitivity of $\alpha \sim 6.3\ten{-11}$ for $\lambda\sim
3.9\ten 6$ m (half the orbital radius).

\section{A possible effect of the Earth's radiation belt}

Charged particles, penetrating into the SEE capsule from space and captured
by the test bodies, create electrostatic forces that could substantially
distort the experimental results. Among the sources of such particles one
should mention (i) cosmic-ray showers, (ii) solar flares and (iii) the
Earth's radiation belts (Van Allen belts). The effect of cosmic-ray showers
was estimated in Ref.\,\cite{SD92} and shown to be negligible. Solar flares
are more or less rare events and, although they create very significant
charged particle fluxes, sometimes even exceeding those in the most dense
regions of the radiation belts, one can assume that the SEE measurements
(except those of $\dot G$) are stopped for the period of an intense flare.
On the contrary, the effect of the Van Allen belts is permanent as long as
the satellite orbit passes, at least partially, inside them.

We will show here that the charging is unacceptably high at otherwise
favorable satellite orbits, so that some kind of charge removal technique is
necessary, but this problem may be addressed rather easily by presently
available technology.

The range of the most favorable SEE orbital altitudes, roughly 1400 to 3300
km \cite{SD92}, coincides with the inner region of the so-called inner
radiation belt [21--24],
situated presumably near the plane of the
magnetic equator. This region is characterized by a considerable flux of
high-energy protons and electrons. For a SEE satellite at altitudes near
1500 km the duration of the charging periods is about 12 minutes. Maximum
charging rates occur in the central Atlantic. It should be noted that the
South Atlantic Anomaly (SAA) --- a region of intense Van Allen activity
which results from the low altitude of the Earth's magnetic field lines over
the South Atlantic Ocean --- cannot cause additional problems for the SEE
experiments. The reason is that the SAA mostly contains low-energy protons
which cannot penetrate into the SEE capsule.

Electrons are known to be stopped by even a thin metallic shell, so only
protons are able to induce charges on the test bodies. Proton-induced
charges on the test bodies can create considerable forces. The inner
radiation belt contains protons with energies of 20 to 800 MeV, and their
maximum fluxes at an altitude of 3000 km over the equator are as great as
about $3\ten 6 \mbox{${\rm cm}^{-2}{\rm s}^{-1}$}\ $ for energies
$E \gsim 10^6$ eV and about $2\ten 4 \mbox{${\rm cm}^{-2}{\rm s}^{-1}$}
\ $ for $E \gsim 10^7$ eV. At 1500 km altitude these numbers are a few times
smaller; the fluxes gradually decrease with growing latitude $\varphi$ and
actually vanish at $\varphi\sim 40\deg$.

It is important that estimates of any resulting effects take into account
that (i) the capsule walls have a considerable thickness and stop the
low-energy part of the proton flux and (ii) among the protons that
penetrate the capsule and hit the Particle, the most energetic ones, whose
path in the Particle material is longer than the Particle diameter, fly it
through and hit the capsule wall again. As for the Shepherd, its size is
large enough to stop the overwhelming majority of protons which hit it.

In what follows, we will assume a Shepherd radius of 20 cm and a Particle
radius of 2 cm and estimate the captured charges for some satellite orbits
in a capsule whose walls of aluminium are 2, 4, 6 and 8 cm thick. The SEE
satellite must actually involve several coaxial cylinders for
thermal-radiation control, and the combined thickness of their walls must
amount to several cm. We will assume, in addition, that the Particle also
consists of aluminium and stops all protons whose path is shorter than 4 cm
(thus overestimating the charge by a small amount since most of protons will
cover a smaller path through the Particle material). A 100 g Particle of
aluminium will have a radius of $\approx 2.07$ cm.

It is advisable to determine first which charges (and fluxes that create
them) might be neglected.

\subsection{Admissible charges}

Let us estimate the Coulomb interaction both between the Shepherd and the
Particle and between each test body and its image in the capsule walls. To
estimate the spurious effects on the Particle trajectory, it is reasonable
to calculate its possible displacements due to the Coulomb forces from the
growing number of captured charged particles. We assume that the test
bodies are discharged by grounding to the capsule before launching the
Particle in each given experiment.

\medskip\noi
{\bf Criterion.} We will call the induced charges, or the fields they
create, {\sl admissible\/} if they cause a displacement of the Particle with
respect to the Shepherd smaller than a prescribed coordinate measurement
error $\delta l$ (we take here $\delta l = 10^{-6}$ m) for a prescribed
measurement time (we take $t \geq 10^4$ s).

The Coulomb acceleration $a_Q (t) = q_M q_m /(r^2 m)$ (in the Gaussian
system of units) depends on the Shepherd-Particle separation $r$ and on the
form of the function $J(t)$, which in turn depends on the satellite orbital
motion.

The charge-induced Particle displacement is approximately
\beq
	\Delta l = \int dt \biggl[\int dt\,a_Q(t)\biggr]
\eeq
since the acceleration is almost unidirectional. If, for estimation
purposes, we suppose that the flux $J(t,x)$ is time-independent, then the
resulting displacement is about
\beq
	\Delta l \sim \frac{1}{30}
	\frac{e^2 S_M S_m J_0^2 t^4}{r^2 m},\label{B5}
\eeq
where $S_M \approx 1256$ cm$^2$ is the Shepherd cross-section, $S_m
\approx 12.56$ cm$^2$ is the Particle cross-section, $m$ is the Particle
mass and $r$ is an average Shepherd-Particle separation.

The strong time dependence is explained by the rapid growth of the Coulomb
force due to growing charges. Numerically, with the above values of $S_M$
and $S_m$, taking $m=100$ g and $r=1$ m (the latter leads to an
overestimated force since the Particle spends most of time at greater
distances), we find:
\beq
	J_0^2 t^4 \lsim  0.83\ten{18}\ {\rm s}^2 {\rm cm}^{-4}. \label{B6}
\eeq
For $t= 10^4$ s an admissible flux is thus less than 9
\mbox{${\rm cm}^{-2}{\rm s}^{-1}$}.

Another undesired effect is that the Particle, being charged by the belt
protons, will interact with the capsule walls. This is well approximated as
an interaction with the Particle's mirror image in the wall, while the
latter may be roughly imagined as a conducting plane. Then, assuming that
the Particle is at average at about 25 cm from the capsule wall and using
the same kind of reasoning as above, we obtain instead of (\ref{B6})
\beq
	J_0^2 t^4 \lsim
		2.07\ten{19}\ {\rm s}^2 {\rm cm}^{-4} \label{B7}
\eeq
and an admissible proton flux less than 45 \mbox{${\rm cm}^{-2}{\rm
s}^{-1}$}\ for $t=10^4$ s.

Some more estimates are of interest: if the charge can be kept smaller
than a certain value, then what is the upper limit for it to create only
negligible displacements? Suppose that there are constant charges on both
the Shepherd ($q=q_M$) and the Particle ($q=q_m,\ m=100$ g), then they are
admissible according to the above criterion as long as
\bear
q_M q_m \lal <2\ten{-6}\,\mbox{${\rm CGSE}_q$}^2 = \fract{2}{9}\ten{-24}\
{\rm C}^2, \label{B8}\\ q_m^2 \lal <\half\ten{-6}\,\mbox{${\rm CGSE}_q$}^2.
\label{B9}
\ear
These inequalities follow, respectively, from considering the
Shepherd-Particle interaction and the interaction between the Particle
(located at 25 cm from the wall) and its image. Thus the maximum admissible
Particle charge is about $7\ten{-4}\,\mbox{${\rm CGSE}_q$} \approx 1.5\ten{6}
e$; assuming this value, it follows from (\ref{B8}) that the maximum
Shepherd charge is about $3\ten{-3}\,\mbox{${\rm CGSE}_q$} \approx 5.5
\ten{6} e$. With these charge values the electric potentials on the test
body surfaces are
\bearr
   U_M \approx 1.5\ten{-4}\,\mbox{${\rm CGSE}_q$}/{\rm cm} = 45\ {\rm mV};
\nnn
   U_m \approx 3.5\ten{-4}\,\mbox{${\rm CGSE}_q$}/{\rm cm} = 105\ {\rm mV}.
\label{b10}
\ear

If by any means the requirements (\ref{B8}), (\ref{B9}) are satisfied (e.g.,
the potentials are kept smaller than the values (\ref{b10})), the
electrostatic effect on the Particle trajectory may be neglected.

\begin{table*}
{\bf Table 2.}
Average flux, peak flux and captured charges per revolution in some
satellite orbits

\begin{center}
\begin{tabular}{|c|c|c|c|r|r|}
\hline
Orbit  &  Wall     &  Average flux,& Peak    flux,& Shepherd \z      &    Particle \z  \\
       & thickness &    \flun\z    &   \flun\z    & charge $q_M$     &    charge $q_m$ \\
\hline
       &  2 cm     &  \nhq 1420    & \nhq 12300  & 1.5\ten{10} $e$  &   4.5\ten 7 $e$ \\
1500b  &  4 cm     &  \nhq 1000    &      8800   &   1\ten{10} $e$  &   2.6\ten 7 $e$ \\
       &  6 cm     &   770         & 6800        & 7.5\ten 9 $e$    &   1.5\ten 7 $e$ \\
       &  8 cm     &   600         & 5400        &   6\ten 9 $e$    &   1.2\ten 7 $e$ \\
\hline
       &  2 cm     &   646         & 5700        & 6.5\ten 9 $e$    &   1.9\ten 7 $e$ \\
1500c  &  4 cm     &   464         & 4200        & 4.3\ten 9 $e$    &   1.1\ten 7 $e$ \\
       &  6 cm     &   365         & 3300        & 3.5\ten 9 $e$    &     7\ten 6 $e$ \\
       &  8 cm     &   280         & 2700        & 2.7\ten 9 $e$    &     5\ten 6 $e$ \\
\hline
\end{tabular}
\end{center}

\end{table*}
The Shepherd's interaction with the respective image charge induced at
nearest location to it in the SEE experimental chamber does not lead to
appreciable Particle displacements. A very demanding requirements on the
Shepherd's charge emerges, however, if the SEE satellite is used for G-dot
determination (whose detailed discussion is postponed to future papers).
For that case,

\beq
	U_M  \lsim 1 \ {\rm mV}.
\eeq
Evidently, in this case the Shepherd-Particle interaction {\it per se\/} is
not the determining factor with respect to charge limits on the test bodies.

\subsection{The captured charges in certain orbits}

The charges captured by the Shepherd and the Particle on board a satellite
in various circular orbits for a single revolution around the Earth, a
period of about two hours, were estimated in~\cite{gracos}. (Actual
measurement times may exceed this period, but not by much.) These estimates
were obtained with the aid of SEE2 and SEREIS software, created at
Nuclear Physics Institute of Moscow State University \cite{niiyaf}.

The results obtained lead to some conclusions of importance for the SEE
experiments (details see in~\cite{gracos}).

First, the models show zero proton fluxes in equatorial orbits of 500 ---
800 km altitude but indicate considerable fluxes at the same altitudes due
to crossing the SAA. It turns out, however, that the SAA is overwhelmingly a
low-energy phenomenon and leaves the fluxes virtually unaffected on the
relevant energy scale, beginning at approximately 65 MeV. Moreover,
there is a very small proton flux due to the SAA even at energies above 10
MeV, hence, with 1 mm layer of shielding, the SAA influence is
negligible.  Behind a thicker layer of shielding there are actually no
secondary particles due to SAA protons.

Second, at an altitude of 1500 km the fluxes depend substantially on the
orbit orientation but remain on the same scale of a few million protons per
cm$^2$ at energies over 65 MeV.

Third, at an altitude of 3000 km both the total flux (for $x=0$) and
its high-energy component in particular are evidently a few times greater
than at 1500 km.

Fourth, and most important: for all orbits in the desirable range of
altitudes the charges are quite large compared with their admissible
values and they remain large even behind relatively thick walls. It is thus
quite important to have means to detect and remove these charges during the
measurements. Moreover, as seen from the peak values in Table 2 and
time scans of Van Allen charging in orbits of interest (1500c is one of the
most favourable ones, 1500b is less favourable; the data were also obtained
using the above-mentioned software), at a charging peak when crossing the
magnetic equator the time required for the charge on the test bodies to
reach its maximum allowable values, as listed above, is a matter of
seconds, not minutes. Therefore the charge must be detected and removed as
it builds up, on a time scale of seconds.

The detection and measurement of the charge on the test bodies can probably
be achieved relatively easily by an array of minute microvoltmeters attached
to the inner wall of the experimental chamber.

Several methods for removing positive charge are now being evaluated. A
simple and promising method may be to shoot electron beams directly at test
bodies. The number of electrons needed is of the order of $10^8/$s.
Although this approach has the inherent drawback that it requires that an
active system must perform correctly for many years, it is simple in
principle and will accomplish the goal.

\section{Concluding remarks}

The main results of the recent developments described in this paper may be
summarized as follows:

1. Numerical simulations of the Particle relative motion in the case of a
   lighter Shepherd (200 kg), for initial Particle-Shepherd separations
   between 2 and 5 m, for $\Horb = 500 \div 3000$ km, has shown that the
   sensitivity of trajectories with respect to changes of $G$ did not
   change too much compared to the previous estimates with a heavier
   Shepherd (500 kg). In particular, variations $\delta G/G \sim 10^{-6}$
   lead to trajectory shifts along the $x$ axis ranging from
   $0.634\ten{-6}$ to $3.34\ten{-6}$ m.

   It has been found that an error of 1 cm in $\Horb$ leads to trajectory
   shifts of 3\ten{-8} to $10^{-7}$ m, i.e., about an order of magnitude
   smaller than the planned coordinate measurement error, $10^{-6}$ m,
   and than the trajectory shifts due to $G$ variations of 1 ppm.

   An estimated error $\delta G/G$ due to Shepherd non-sphericity,
   characterized by its quadrupole moment $J_2 \sim 10^{-5}$, is close to
   $10^{-6}$; this implies rather severe requirements upon the Shepherd
   fabrication precision. The same quadrupole moment causes even
   greater $\delta G/G$ if the Particle motion starts at $y_0 > 15$ cm.

2. Computer simulations have shown that the gravitational constant $G$ can
   be measured up to $\sim 5\ten{8}$. The inverse square law may be
   tested with a sensitivity of $\alpha\sim 2\ten {-7}$ for $\lambda = 1.2$
   m and $\alpha \sim 3\ten{-8}$ for $\lambda\sim 3.4\ten 6$ m (half the
   Earth's radius). Observation of the Shepherd orbit precession makes it
   possible to test the inverse square law at $\lambda \sim 3.4\ten 5$ m up
   to $10^{-10}$.

3. Estimation of test body charging due to crossing the van Allen radiation
   belts shows that this effect requires special means for charge
   measuring and removal. These means, however, do not go beyond the
   presently available technology level.

\Acknow
{This work was supported in part by NASA grant \# NAG 8-1442. K.A.B. wishes
to thank Nikolai V. Kuznetsov for helpful discussions and for providing
access to the SEE2 and SEREIS software. V.N.M. is grateful to CINVESTAV and
CONACYT for their hospitality and support during his stay in Mexico.  }

\end{document}